\documentclass[letterpaper,11pt]{article}
\usepackage{amssymb}
\usepackage{amsmath}
\usepackage{amstext}
\usepackage{graphicx,epsfig}
\usepackage{epsfig}
\usepackage{verbatim} 
\usepackage{fancybox}
\usepackage{color}
\usepackage{ulem}
\usepackage{enumitem}
\usepackage{subfigure}
\usepackage{bbm}
\usepackage{parskip}
\usepackage{dsfont}
\usepackage[numbers,sort&compress]{natbib}
\usepackage[body={6.5in, 9in}, right=1in, top=1in]{geometry}

\newcommand{\be}{\begin{equation}}
\newcommand{\ee}{\end{equation}}
\newcommand{\bea}{\begin{eqnarray}}
\newcommand{\eea}{\end{eqnarray}}
\newcommand{\beas}{\begin{eqnarray*}}
\newcommand{\eeas}{\end{eqnarray*}}

\def\({\left(}
\def\){\right)}

\def\gsim{ \lower .75ex \hbox{$\sim$} \llap{\raise .27ex \hbox{$>$}} }
\def\lsim{ \lower .75ex \hbox{$\sim$} \llap{\raise .27ex \hbox{$<$}} }

\usepackage{xcolor,colortbl}

\begin{document}
%\maketitle

\begin{center}
\Large{\textbf{Chameleon Dark Energy and Atom Interferometry}} \\[0.5cm]
 
\large{Benjamin Elder$^1$, Justin Khoury$^1$, Philipp Haslinger$^3$, Matt Jaffe$^3$, Holger~M\"{u}ller$^{3,4}$, \\ Paul Hamilton$^2$}
\\[0.5cm]

\small{
\textit{$^1$Center for Particle Cosmology, Department of Physics and Astronomy, \\ University of Pennsylvania, Philadelphia, PA 19104}

%\vspace{.2cm}

\textit{$^2$Department of Physics and Astronomy, University of California, Los Angeles, CA 90095}

\textit{$^3$Department of Physics,  University of California, Berkeley, CA 94720}

%\vspace{.2cm}

\textit{$^4$Lawrence Berkeley National Laboratory, Berkeley, CA, 94720}}

\vspace{.2cm}

\end{center}

%\tableofcontents

\vspace{.2cm}

\hrule \vspace{0.2cm}
\centerline{\small{\bf Abstract}}
{\small Atom interferometry experiments are searching for evidence of chameleon scalar fields with ever-increasing precision.  As experiments become more precise, so too must theoretical predictions.  Previous work has made numerous approximations to simplify the calculation, which in general requires solving a 3-dimensional nonlinear partial differential equation (PDE).  In this paper, we introduce a new technique for calculating the chameleonic force, using a numerical relaxation scheme on a uniform grid.  This technique is more general than previous work, which assumed spherical symmetry to reduce the PDE to a 1-dimensional ordinary differential equation (ODE).  We examine the effects of approximations made in previous efforts on this subject, and calculate the chameleonic force in a set-up that closely mimics the recent experiment of Hamilton {\it et al.} Specifically, we simulate the vacuum chamber as a cylinder with dimensions matching those of the experiment, taking into account the backreaction of the source mass, its offset from the center, and the effects of the chamber walls. Remarkably, the acceleration on a test atomic particle is found to differ by only 20\% from the approximate analytical treatment. These results allow us to place rigorous constraints on the parameter space of chameleon field theories, although ultimately the constraint we find is the same as the one we reported in Hamilton {\it et al.} because 
we had slightly underestimated the size of the vacuum chamber. This new computational technique will continue to be useful as experiments become even more precise, and will also be a valuable tool in optimizing future searches for chameleon fields and related theories.}   
\vspace{0.3cm}
\noindent
\hrule
\def\thefootnote{\arabic{footnote}}
\setcounter{footnote}{0}

\section{Introduction}

Over the past decade there has been tremendous activity, both theoretical and experimental, devoted to theories of the dark sector with new light degrees of freedom that
couple to ordinary matter and mediate a fifth force~\cite{Joyce:2014kja}. These degrees of freedom (generally considered to be scalar fields) couple to matter with strength comparable to, or stronger than, the gravitational force. Nevertheless they have managed to escape detection (thus far) through so-called {\it screening mechanisms}. In regions of high density, where experiments are performed, the scalar fields develop strong non-linearities which result in an effective decoupling and correspondingly weak force. Thus screening mechanisms
rely on the interplay between the interactions with matter and the non-linear self-interactions of the scalar. 

Broadly speaking, one distinguishes two universality classes of screening mechanisms:

\begin{itemize}

\item In the first universality class, scalar non-linearities arise from a self-interaction potential $V(\phi)$. As a result, whether a source is screened or not depends
on the local scalar field value. This class includes chameleons~\cite{Khoury:2003aq,Khoury:2003rn,Gubser:2004uf,Brax:2004qh,Mota:2006ed,Mota:2006fz},
symmetrons~\cite{Hinterbichler:2010es,Olive:2007aj,Pietroni:2005pv,Hinterbichler:2011ca,Brax:2011pk}, varying-dilatons~\cite{Brax:2011ja} and their variants. 

\item In the second universality class, scalar non-linearities arise from derivative interactions. In this case, whether a source is screened or not depends on
the local field gradients. This class includes K-mouflage-type scalars~\cite{Babichev:2009ee,Dvali:2010jz,Burrage:2014uwa}, galileons~\cite{Deffayet:2001uk,Luty:2003vm,Nicolis:2004qq,Nicolis:2008in}
and disformally-coupled scalars~\cite{Bekenstein:1992pj,Zumalacarregui:2010wj}.

\end{itemize}

Theories within the same universality class all lead to similar phenomenology, even though their Lagrangians may look quite different. Theories belonging to the first universality class yield the richest phenomenology on small scales, including in the laboratory and in the solar system. On the other hand, the range of the scalar-mediated force
can be at most $\sim {\rm Mpc}$ cosmologically~\cite{Wang:2012kj,Brax:2011aw}. Theories belonging to the second universality class have the largest impact on scales larger than $\sim {\rm Mpc}$, but presently lead to unmeasurably small effects in the laboratory~\cite{Brax:2011sv}. 

In this paper we focus on chameleon scalar field theories, though our methods and results can be generalized to other theories in the first universality class. In chameleon theories, the mass of chameleon particles depends on the local environmental matter density, which is the result of an interplay between their self-interaction potential and their coupling to ordinary matter. In dense regions, such as in the laboratory, the mass of the chameleon is large, and the resulting force mediated by the chameleon is short-ranged, shielding the chameleon interaction from detection. In regions of low density, such as in space, the mass of chameleon particles can be much smaller, and the resulting force mediated by the chameleon is long-ranged. 

The simplest Lagrangian for a chameleon theory is
\begin{equation}
L_{\rm cham} =  -\frac{1}{2}(\partial \phi)^2 - V(\phi) -  \frac{\phi}{M} \rho_{\rm m} \,,
\label{actionintro}
\end{equation}
where $\rho_{\rm m}$ is the matter density, assumed to be non-relativistic. The chameleon mechanism is achieved for various different potentials. For concreteness, in this paper we will focus on the inverse power-law form~\cite{Ratra:1987rm,Wetterich:1987fm}
\begin{equation}
V(\phi) = \Lambda^4 \left(1 + \frac{\Lambda^n}{\phi^n} \right)\,;\qquad n > 0\,.
\label{trackerpotintro}
\end{equation}
The inverse power-law form, considered in the original chameleon papers~\cite{Khoury:2003aq,Khoury:2003rn}, is motivated by earlier studies of 
tracker quintessence models~\cite{Zlatev:1998tr, Steinhardt:1999nw} and arises generically from non-perturbative effects in supergravity/string theories, {\it e.g.},~\cite{Binetruy:1996nx,Barreiro:1997rp,Binetruy:1997vr}.
(Potentials with {\it positive} powers, $V(\phi) \sim \phi^{2s}$ with $s$ an integer $\geq 2$, can also realize the chameleon mechanism~\cite{Gubser:2004uf}.) 
The constant piece can drive cosmic acceleration at the present time for $\Lambda = \Lambda_0 \simeq 2.4$~meV. The $1/\phi^n$ term is responsible for the non-linear scalar interactions required for the chameleon mechanism to be operational. 

This is a specific example of a fifth force being sensitive to its environment, an idea which has spurred a great deal of activity. Astrophysically, chameleon scalars affect the
internal dynamics~\cite{Hui:2009kc,Jain:2011ji} and stellar evolution~\cite{Chang:2010xh,Davis:2011qf,Jain:2012tn} of dwarf galaxies residing in voids or
mildly overdense regions. In the laboratory, chameleons have motivated multiple experimental efforts aimed at searching for their signatures, including
torsion-balance experiments~\cite{Kapner:2006si,Upadhye:2012qu}, Bose-Einstein condensates~\cite{Harber:2005ic}, gravity resonance spectroscopy~\cite{Ivanov:2012cb,Jenke:2014yel} and neutron interferometry~\cite{Brax:2011hb,Brax:2013cfa,Lemmel:2015kwa,Li:2016tux}. Assuming an additional coupling between photons and chameleons, the CHameleon Afterglow SEarch (CHASE) experiment~\cite{Chou:2008gr,Steffen:2010ze} has looked for an afterglow from trapped chameleons converting into photons. Similarly, the Axion Dark Matter eXperiment (ADMX) resonant microwave cavity was used to search for chameleons~\cite{Rybka:2010ah}. Photon-chameleon mixing can occur deep inside the Sun~\cite{Brax:2010xq} and affect the spectrum of distant astrophysical objects~\cite{Burrage:2008ii}.
 
Our primary interest is atom interferometry. Following the initial theory papers promoting this
method~\cite{Burrage:2014oza,Burrage:2015lya}, we carried out an experiment at UC Berkeley to search and constrain the chameleon parameter
space~\cite{Hamilton:2015zga}. The experiment measures the force between an aluminum sphere (the ``source" mass) and ${}^{133}$Cs atoms (the ``test" mass). Because the experiment  is performed in vacuum, the chameleon Compton wavelength is comparable to the size of the vacuum chamber and hence relatively long-ranged on the scale of the
experiment. Moreover, due to their microscopic size, the Cs atoms are unscreened and hence act as test particles. The chameleon force they experience is still
suppressed by the fact that the source mass is screened, but less so than the force between two macroscopic objects. With this set-up, we were
able to bound an anomalous contribution to the acceleration: $a < 5.5 ~ \mu \mathrm{m} / \mathrm{s}^2$ at the 95\% confidence level~\cite{Hamilton:2015zga}. 

To translate this into a constraint on the chameleon parameter space, in~\cite{Hamilton:2015zga} we used a number of analytical approximations. Specifically, we treated
the vacuum chamber as a sphere and ignored the details of chamber walls. The assumption of spherical symmetry reduces the static equation of motion,
which is a three-dimensional partial differential equation (PDE), to a one-dimensional ordinary differential equation (ODE) that can easily be integrated numerically.
We then calculated the force between source mass and atoms using approximate analytical expressions derived in the early chameleon papers~\cite{Khoury:2003aq,Khoury:2003rn}. In the past these methods have proven to do a fairly good job at estimating the chameleon profile in various situations. But if we are to rigorously exclude part of the chameleon theory space, a more accurate treatment is warranted.

In this paper we present a scheme to solve the full three-dimensional PDE for the chameleon profile in the vacuum chamber, making it possible to calculate the force
due to the chameleon field at any point and along any direction. This allows us to relax the assumption of spherical symmetry, and to therefore accurately model the
cylindrical vacuum chamber used in~\cite{Hamilton:2015zga}. Furthermore, we can exactly and consistently include the effects of the chamber walls and the source mass,
which is offset from the center, without having to resort to approximate analytical expressions. 

The motivations for this work are three-fold. Firstly, the exact approach followed here allows us to quantify the validity of the approximations made in~\cite{Hamilton:2015zga},
as well to place rigorous constraints on chameleon theories from the experimental bound on $a$. Secondly, it allows us to check claims in the literature
that accounting for the chamber walls leads to a significant effect on the field profile deep inside the chamber~\cite{Schlogel:2015uea} or that the thin-shell expression that goes
back to~\cite{Khoury:2003aq,Khoury:2003rn} gives a poor approximation to the chameleon force~\cite{Kraiselburd:2015vyf}. We will see that these claims are wrong. A detailed treatment of the walls has negligible effect inside the chamber, a conclusion that is now shared by the authors of~\cite{Schlogel:2015uea} in a revised version of their paper. We will also find that the thin-shell approximation works remarkably well.

Our main findings are at once reassuring and disappointing! The analytical approximations made in~\cite{Hamilton:2015zga} work remarkably well and unexpectedly well. Specifically, carefully simulating the vacuum chamber as a cylinder with dimensions matching those of~\cite{Hamilton:2015zga}, taking into account the backreaction of the source mass, its offset from the center, and the effects of the chamber walls, the acceleration on a test atomic particle is found to differ by only 20\% from the simplified analysis of~\cite{Hamilton:2015zga}. A 20\% difference would be barely visible on the logarithmic exclusion plots,
but the actual difference is even smaller, thanks to a fortuitous cancellation. Namely, while the acceleration in~\cite{Hamilton:2015zga} is a slight overestimate (by $\sim 20\%$) of the actual answer, this is compensated by an a slight {\it underestimate} of the vacuum chamber radius (5~cm instead of the actual 6~cm). These two ``mistakes" interfere destructively, leaving us with almost identical constraints
on the chameleon parameters. We apologize to the reader for the lack of drama. Being that most of us were authors on~\cite{Hamilton:2015zga}, we view this outcome as
quite positive.

Looking ahead, our code can be used to determine the ideal source mass geometry and position to optimize the chameleon signal in future experiments. Although our
treatment is cast in the context of atom interferometry, the code is quite versatile and can be applied to {\it any} experiment --- atom interferometry, cold neutrons or a torsion pendulum --- aimed at constraining the chameleon field inside a vacuum chamber.  To illustrate the usefulness of the code, we will apply it in Sec.~\ref{sec:forecasts} to 
forecast the signal in an improved version of our experiment, as well as for a proposed interferometry experiment to take place in NASA's Cold Atom Laboratory~\cite{CAL} aboard the International Space Station.

This paper is organized as follows. We give a brief review of the chameleon mechanism in Sec.~\ref{sec:model}, including a discussion of the thin-shell approximate treatment
used in~\cite{Hamilton:2015zga}. We summarize existing experimental constraints and motivations for the present work in Sec.~\ref{sec:motivations}. After a brief description
of our numerical method in Sec.~\ref{sec:numerical_method}, we present the results of 3D integration as a series of refinements, from the crude ``spherical cow" approximation made
in~\cite{Hamilton:2015zga} all the way to the actual experimental set-up with cylindrical chamber and offset source mass in Sec.~\ref{sec:results}. In Sec.~\ref{realsim} we simulate the chameleon profile with
the experimental set-up~\cite{Hamilton:2015zga} for a range of chameleon parameters, and derive realistic constraints on the space of chameleon theories.  In Sec.~\ref{sec:forecasts} we report results on ongoing and upcoming experiments.  We summarize our results and discuss future applications in Sec.~\ref{sec:conclusions}.

\begin{figure}[t]
\centering
\includegraphics[width=3.1in]{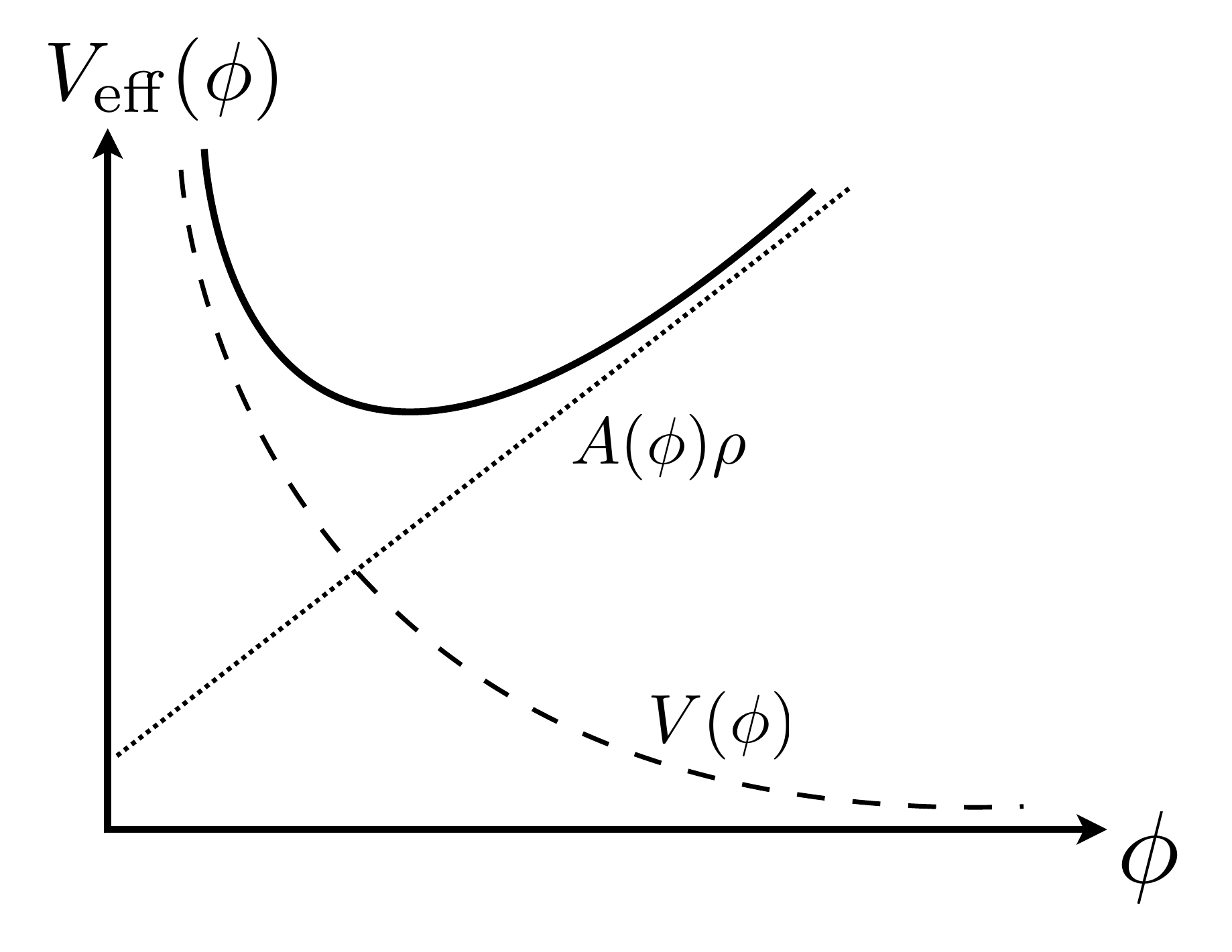}
\caption{\label{chameffpot} \small Schematic of the effective potential felt by a chameleon field (solid line), given by the sum of the bare potential of runaway form, $V(\phi)$ (dashed line), and a density-dependent piece, from coupling to matter (dotted line).}
\end{figure}

\section{Chameleons: A Brief Review}
\label{sec:model}

A chameleon scalar field has the defining property of coupling to matter in such a way that its effective mass increases with increasing local matter density~\cite{Khoury:2003aq,Khoury:2003rn,Gubser:2004uf,Brax:2004qh,Mota:2006ed,Mota:2006fz}. The scalar-mediated force between matter particles can be of gravitational strength (or even stronger), but its range is a decreasing function of ambient matter density, 
and therefore avoids detection in regions of high density. Deep in space, where the mass density is low, the scalar is light and mediates a fifth force of gravitational strength, but near the Earth, where experiments are performed, and where the local density is high, it acquires a large mass, making its effects short ranged and hence unobservable. 

\subsection{Theoretical set up}

In the Newtonian limit where matter is non-relativistic, the Lagrangian for a prototypical chameleon theory is
\begin{equation}
L_{\rm cham} =  -\frac{1}{2}(\partial \phi)^2 - V(\phi) -  A(\phi) \rho_{\rm m} \,.
\label{action}
\end{equation}
This generalizes~\eqref{actionintro} to include a general coupling function $A(\phi)$ to the matter density $\rho_{\rm m}$. For simplicity, we assume that the chameleon scalar field $\phi$ couples universally to matter, {\it i.e.}, via a single function $A(\phi)$. Generalizations involving different coupling functions for different matter species are also possible, resulting in violations of the weak equivalence principle. In the simpler case of interest, the theory is characterized by two functions: the self-interaction potential $V(\phi)$ and the coupling function to matter $A(\phi)$. The coupling function is assumed to be approximately linear,\footnote{In the symmetron~\cite{Hinterbichler:2010es,Olive:2007aj,Pietroni:2005pv,Hinterbichler:2011ca,Brax:2011pk} and varying-dilaton~\cite{Brax:2011ja} mechanisms, on the other hand, a $\phi\rightarrow -\phi$ symmetry precludes the linear term in $A(\phi)$. The appropriate form in those classes of theories is $A(\phi) \simeq 1 + \frac{\phi^2}{M^2}$. In practice, however, the phenomenology of symmetrons/varying-dilatons is qualitatively similar to that of the chameleon.}
\be
A(\phi) \simeq 1 + \frac{\phi}{M}\,.
\label{Aphi}
\end{equation}
To compare with experiments we will be primarily interested in the range $10^{-5}\;M_{\rm Pl}\;\lsim\; M \;\lsim\; M_{\rm Pl}$,
where $M_{\rm Pl} = (8 \pi G_{\rm N})^{-1/2} \simeq 2.4 \times 10^{18}$~GeV is the reduced Planck mass. This range of $M$ is interesting because it has not yet been experimentally ruled out. Over this range, the field excursion is much smaller than $M$ throughout the apparatus, and hence the linear approximation~\eqref{Aphi} is justified.

For the self-interaction potential, as mentioned in the Introduction we specialize to the Ratra--Peebles inverse power-law form~\cite{Ratra:1987rm,Wetterich:1987fm}
\begin{equation}
V(\phi) = \Lambda^4 \left(1 + \frac{\Lambda^n}{\phi^n} \right)\,,
\label{trackerpot}
\end{equation}
with $n > 0$. The constant piece can drive cosmic acceleration at the present time for $\Lambda = 2.4$~meV, whereas the $1/\phi^n$ piece is responsible for the chameleon mechanism. 

\begin{figure}[t]
\centering
\includegraphics[width=5.1in]{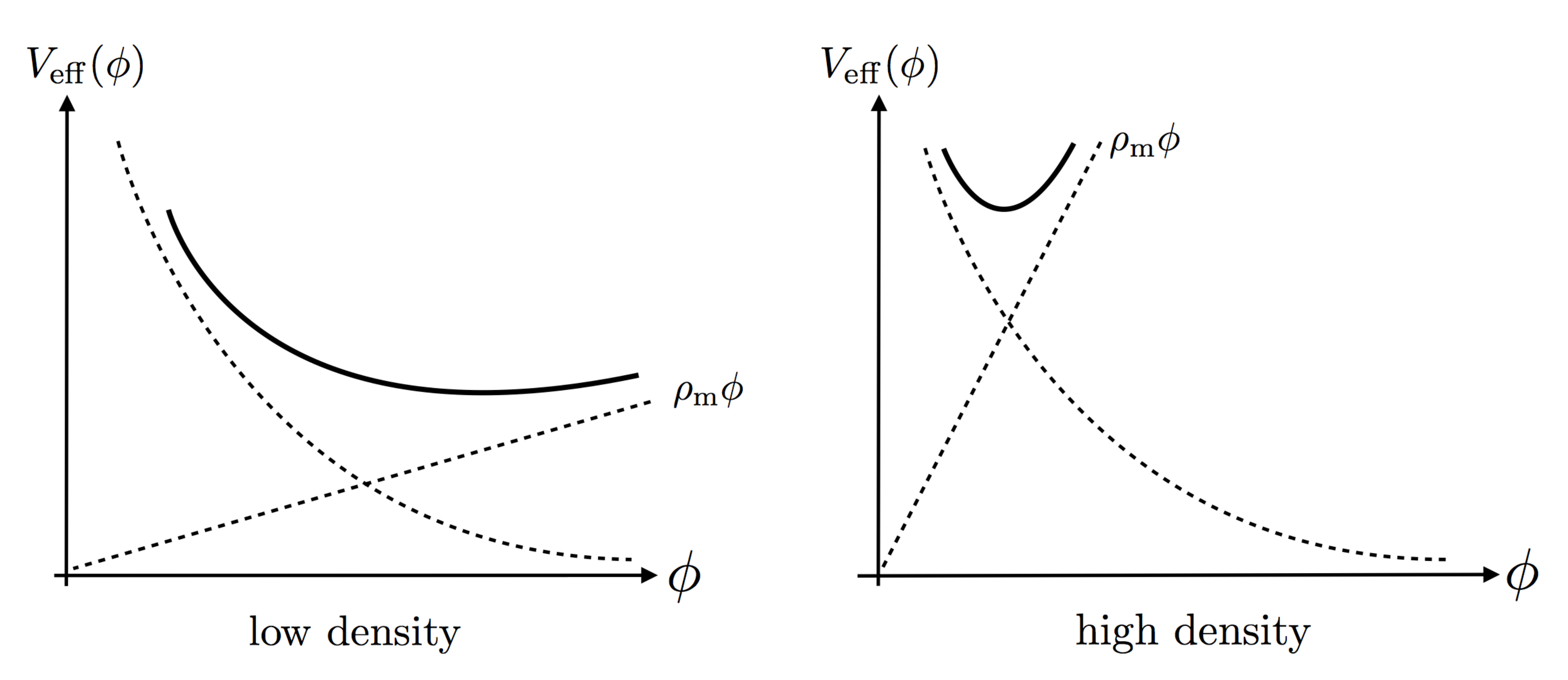}
\caption{\label{champotcomp} \small Effective potential for low ambient matter density (Left) and high ambient density (Right). As the density increases, the minimum of the effective potential, $\phi_{\rm min}$, shifts to smaller values, while the mass of small fluctuations, $m_\phi$, increases.}
\end{figure}

It is clear from the action~\eqref{action} that the scalar field is governed by a density-dependent effective potential
\be
V_{\rm eff}(\phi) = V(\phi) + A(\phi) \rho_{\rm m} \,.
\label{Veffcham}
\ee
This is sketched in Fig.~\ref{chameffpot}. In an environment of homogeneous $\rho_{\rm m}$, the effective potential is minimized at
\be
\phi_\mathrm{eq} = \left( \frac{ n M  \Lambda^{4 + n}}{\rho_{\rm m}} \right)^{\frac{1}{n + 1}}~.
\ee
The mass of chameleon particles around this state, defined as usual by  $m^2(\phi_\mathrm{eq}) = \left.\frac{\partial^2 V_\mathrm{eff}}{\partial \phi^2} \right\vert_{\phi = \phi_\mathrm{eq}}$, is
\be
m_{\rm eq}^2 = \frac{n (n + 1) \Lambda^{4 + n}}{\phi_\mathrm{eq}^{n + 2}} \sim \rho_{\rm m}^{\frac{n+2}{n+1}}\,.
\label{mass}
\ee
As the value of $\rho_{\rm m}$ increases, we see that $\phi_\mathrm{eq}$ decreases while $m_{\rm eq}$ increases, as desired. This is sketched in Fig.~\ref{champotcomp}.

More generally, to compute the chameleonic acceleration on a test particle due to an arbitrary static distribution of matter, we begin by solving for the $\phi$ profile:
\be
\vec \nabla^2 \phi = V_\mathrm{eff},_\phi~,
\label{eom}
\ee
For general $\rho_{\rm m}(\vec{x})$, we must of course resort to numerical integration. Given the resulting field profile $\phi(\vec{x})$, the acceleration on a test particle due to the chameleon interaction readily follows from~\eqref{action}:
\be
\vec a = \frac{1}{M} \vec \nabla \phi~.
\label{acceleration}
\ee
For the parameters of interest, we will see that the atoms in the experiment behave as test particles to an excellent approximation. Indeed, this is one of the virtues of using atom-interferometry
to test chameleons! More generally, the chameleon force on an extended body can be computed borrowing a method developed by Einstein, Infeld and Hoffmann~\cite{Einstein:1938yz} in the context of General Relativity, as nicely shown in~\cite{Hui:2009kc}.

\subsection{Thin-shell approximate treatment}

Before solving the chameleon equation of motion exactly using numerical integration, it is helpful to gain intuition on how the chameleon force is suppressed in 
the presence of high ambient density by reviewing the approximate solution first presented in~\cite{Khoury:2003aq,Khoury:2003rn}.
One of the main goals of this paper is to assess to what extent the approximate treatment works. 

Consider a static, spherical source with radius $R$ and homogeneous density $\rho_{\rm obj}$. For the moment, we imagine that this object is immersed in a homogeneous medium with density 
$\rho_{\rm bg}$. (We will come back shortly to the case of the vacuum chamber, where the ambient density is approximately zero.) We denote by $\phi_{\rm obj}$ and $\phi_{\rm bg}$ the minima of the effective potential at the object and ambient density, respectively. The scalar equation of motion reduces to
\begin{equation}
\phi''+\frac{2}{r}\phi' = V_{,\phi}  + \frac{\rho_{\rm m}(r)}{M}~\,; \qquad \qquad \rho_{\rm m}(r) = \left\{\begin{array}{l}
\rho_{\rm obj}~~~~~~~~r < R
\\
\rho_{\rm bg}~~~~~~~~~r > R
\end{array}\right.~.
\label{chameombody}
\end{equation}
The boundary conditions are $\phi'(r=0) = 0$, enforcing regularity at the origin; and $\phi\rightarrow \phi_{\rm bg}$ as $r\to \infty$,
which minimizes the effective potential far from the source.

For a sufficiently large body --- in a sense that will be made precise shortly --- the field approaches the minimum of its effective potential deep in its interior:
\be
\phi \simeq \phi_{\rm obj}~;~~~~~~~~~~~~r < R \,.
\ee
Outside of the object, but still within an ambient Compton wavelength away ($ r < m^{-1}_{\rm bg}$), the field profile goes approximately as $1/r$:
$\phi \simeq \frac{C}{r} +  \phi_{\rm bg}$. One integration constant has already been set to fulfill the second boundary condition above. The other constant $C$
is fixed by matching the field value at $r=R$, with the result
\begin{equation}
\phi \simeq - \frac{R}{r}(\phi_{\rm bg} - \phi_{\rm obj}) + \phi_{\rm bg}\,.
\label{phiext}
\ee

Further intuition on this solution follows from a nice analogy with electrostatics~\cite{JonesSmith:2011tn,Pourhasan:2011sm}. Since $\nabla^2 \phi\simeq 0$ both inside and outside the source, the body acts as a chameleon analogue of a conducting sphere. Any chameleon ``charge" is confined to a thin shell of thickness $\Delta R$ near the surface. The surface ``charge density" $\frac{\rho_{\rm obj} \Delta R}{M}$ must support the discontinuity in field gradients:
\begin{equation}
\left.\frac{{\rm d} \phi}{{\rm d}r}\right\vert_{r=R^+} = \frac{\rho_{\rm obj}\Delta R}{M} \,.
\end{equation}
Substituting~(\ref{phiext}) fixes the shell thickness:
\begin{equation}
\Delta R  = \frac{M\phi_{\rm bg}}{\rho_{\rm obj}R} \,,
\label{DelR}
\end{equation}
where we have assumed $\phi_{\rm bg} \gg \phi_{\rm obj}$ appropriate for large density contrast. For consistency, we should have $\Delta R  \ll R$, in other words $\frac{M\phi_{\rm bg}}{\rho_{\rm obj}R^2} \ll 1$. In that case the object is said to be {\it screened}. The acceleration on a test particle located within $r < m^{-1}_{\rm bg}$ away is
\be
a \simeq a_{\rm N} \left( \frac{M_\mathrm{Pl}}{M} \right)^2 \frac{6\Delta R}{R} \qquad ({\rm screened})\,,
\label{accel_approx}
\ee
where $a_{\rm N}$ is the Newtonian acceleration. If instead $\frac{M\phi_{\rm bg}}{\rho_{\rm obj}R^2} \gg 1$, the object is said to be {\it unscreened}, and the exterior acceleration is unsuppressed:
\be
a \simeq 2 a_{\rm N} \left( \frac{M_\mathrm{Pl}}{M} \right)^2  \qquad ({\rm unscreened})\,.
\label{accel_approx_unscreened}
\ee

In the case of a vacuum chamber, the background density is so small that the Compton wavelength $m^{-1}_{\rm bg}$ is much larger than the radius of the chamber, hence the field is unable to minimize its effective potential. Instead the scalar field approaches a value about which the Compton wavelength is comparable to the size of the vacuum chamber, $m^{-1}_{\rm vac} \sim R_{\rm vac}$. In other words, from~\eqref{mass} the background value is set by the condition $\phi_\mathrm{vac} \sim \left(n (n + 1) \Lambda^{4 + n} R_{\rm vac}^2 \right)^\frac{1}{n + 2}$. Following~\cite{Hamilton:2015zga} it is convenient to introduce a ``fudge" factor $\xi$, to turn the relation into an equality:
\be
\phi_{\rm bg} = \xi \Big(n (n + 1) \Lambda^{4 + n} R_{\rm vac}^2 \Big)^\frac{1}{n + 2}~.
\label{phi_vac}
\ee
In~\cite{Hamilton:2015zga} it was found that $\xi$ is largely insensitive to $n$, $\Lambda$ and $M$, as well as to the assumed chamber geometry. 
Specifically, for $n=1$ and the dark energy value $\Lambda = 2.4$~meV, one finds $\xi = 0.55$ for a spherical
vacuum chamber and $\xi = 0.68$ for an infinite cylinder. 

The field profile for a spherical source inside a spherical chamber follows identically from the earlier derivation, with $\phi_{\rm bg}$ now given by~\eqref{phi_vac}.
In particular the expression for the shell thickness~\eqref{DelR} becomes $\Delta R  =  \frac{M\xi}{\rho_{\rm obj}R} \left(n (n + 1) \Lambda^{4 + n} R_{\rm vac}^2 \right)^\frac{1}{n + 2}$.
For the parameter values considered here, it is easily seen that the source mass is always screened, {\it i.e.}, the resulting acceleration on a test particle is given by~\eqref{accel_approx}.
Similarly, the atoms are unscreened --- they do not significantly perturb the chameleon field and therefore behave as test particles to an excellent approximation.

\begin{figure}
\centering
\subfigure[]{%
\includegraphics[width = 0.4 \linewidth]{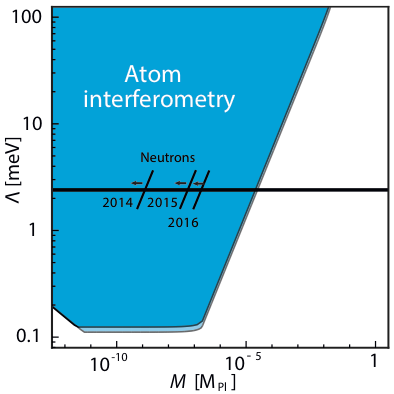}}
\subfigure[]{%
\includegraphics[width = 0.42 \linewidth]{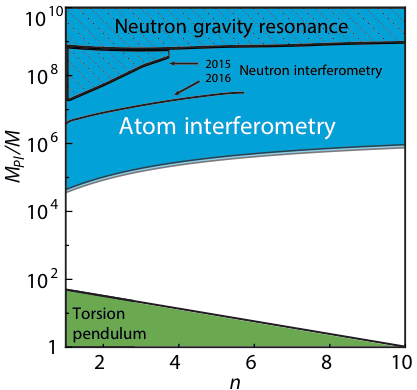}}
\caption{Current constraints due to atom interferometry and torsion pendulum experiments.  We are mainly concerned with $\Lambda = \Lambda_0$, indicated by the black line on the first plot, so that the chameleon field can drive the observed accelerated expansion of the universe. The narrow light blue stripes on the left panel show the influence of varying the fudge parameter over $0.55 \leq \xi \leq 0.68$. The second plot shows $M_\mathrm{Pl}/ M$ {\it vs.} $n$, and also assumes $\Lambda = \Lambda_0$. The ``torsion pendulum" region shown in green has been corrected from~\cite{Hamilton:2015zga} to accurately reflect the constraints imposed by that experiment, following \cite{Upadhye:2012qu}.}
\label{fig:exclusions}
\end{figure}

\section{Existing Constraints and Motivations for this Work}
\label{sec:motivations}

The class of chameleon theories described above are specified by three parameters: the coupling scale $M$, with $M\sim M_{\rm Pl}$ corresponding to gravitational strength fifth force; the scale of the potential $\Lambda$, with $\Lambda = \Lambda_0 \simeq 2.4$~meV corresponding to the value needed to reproduce the observed cosmic acceleration; and the inverse power $n$ specifying the shape of the potential. 

Figure~\ref{fig:exclusions}a) shows current experimental constraints in the $(\Lambda,M)$ plane for $n =1$, where the solid line indicates $\Lambda = \Lambda_0 \simeq 2.4$~meV. The narrow light blue stripes on the left panel show the influence of varying the fudge parameter over $0.55 \leq \xi \leq 0.68$. Meanwhile, Fig.~\ref{fig:exclusions}b) plots the excluded region in the $(M,n)$ plane, with $\Lambda$ fixed to the dark energy value 2.4~meV.  Various experiments contribute to these plots. These include measurements of the Casimir-Polder force using an oscillating ${}^{87}{\rm Rb}$ Bose-Einstein condensate~\cite{Harber:2005ic}, gravity resonance spectroscopy using ultracold neutrons~\cite{Ivanov:2012cb,Jenke:2014yel} and neutron interferometry~\cite{Brax:2011hb,Brax:2013cfa,Lemmel:2015kwa,Li:2016tux}. The E\"ot-Wash torsion balance experiment~\cite{Kapner:2006si} constraint rules out $M \;\gsim\; 10^{-2}~M_{\rm Pl}$ with $\Lambda = \Lambda_0 \simeq 2.4$~meV, corresponding to the lower region of Fig.~\ref{fig:exclusions}b).\footnote{As already mentioned in the Introduction, other experiments constrain the electromagnetic coupling $e^{\beta_\gamma\phi}F_{\mu\nu}F^{\mu\nu}$, which induces photon/chameleon oscillations.} 

In this paper we focus on the Berkeley atom interferometry experiment~\cite{Hamilton:2015zga}, which rules out most of the parameter space shown in the figures. In particular, for $\Lambda = \Lambda_0 \simeq 2.4$~meV and $n=1$ (Fig.~\ref{fig:exclusions}a)) atom interferometry excludes the range $M \;\lsim\; 10^{-5}~M_{\rm Pl}$. The Berkeley experiment, motivated by a theory paper of Burrage {\it et al.}~\cite{Burrage:2014oza}, used atom interferometry to measure the force between ${}^{133}{\rm Cs}$ atoms and an Al sphere. The original experiment constrained an anomalous contribution to the free-fall acceleration as $\Delta a = (0.7\pm 3.7)~\mu {\rm m}/{\rm s}^2$. The excluded regions were then generated using a number of simplifying assumptions:

\begin{itemize}

\item The background chameleon field profile $\phi_{\rm bg}$ was computed $i)$ without the source mass, $ii)$~ignoring the thickness of the chamber walls, and $iii)$~assuming a spherical vacuum chamber. 

\item The chameleon acceleration acting on the atoms was calculated using the thin-shell expression~\eqref{accel_approx} described earlier.

\end{itemize}

The purpose of our paper is to check those assumptions.  We do so by computing the chameleon field profile numerically using a 3-dimensional PDE solver that we developed for this purpose. Our numerical method will be described in detail in the next Section. We solve for the chameleon field profile inside the source sphere, vacuum chamber, and within the vacuum chamber walls.  However, we neglect the backreaction of the atoms, treating them as test particles that do not significantly influence the chameleon field profile.  This assumption is justified by the fact that the atoms are small and light enough to be unscreened for the range of parameters considered here. We will perform a battery of checks, described in detail in Sec.~\ref{sec:results}. 

For the benefit of the anxious reader, we can summarize our findings succinctly as follows: {\it the simplifying assumptions made in~\cite{Hamilton:2015zga} and listed above work remarkably and surprisingly well.} Specifically, carefully simulating the vacuum chamber as a cylinder with dimensions matching those of~\cite{Hamilton:2015zga}, taking into account the backreaction of the source mass, its offset from the center, and the effects of the chamber walls, the acceleration on a test atomic particle is found to differ by only 20\% from the simplified analysis of~\cite{Hamilton:2015zga}. A 20\% difference would be barely visible on a logarithmic scale such as in Fig.~\ref{fig:exclusions}, but the actual difference is even smaller, thanks to a fortuitous cancellation. Namely, while the acceleration~\cite{Hamilton:2015zga} is a slight overestimate (by $\sim 20\%$) of the actual answer, this is compensated by a slight {\it underestimate} of the vacuum chamber radius (5~cm instead of the actual 6~cm). These two ``mistakes" interfere destructively, leaving us with an identical constraint: $M \;\lsim\; 2.3 \times 10^{-5} M_\mathrm{Pl}$ is ruled out for~$\Lambda = \Lambda_0$. 

\section{Numerical Method}
\label{sec:numerical_method}

We integrate the chameleon equation of motion~\eqref{eom} through successive under-relaxation with intermediate steps calculated by the Gauss-Siedel scheme~\cite{NR}. This method is briefly reviewed in the Appendix. We demand that the first derivative of $\phi$ vanish at the edge of the simulation box, which is justified so long as $\phi$ has minimized its effective potential by that point. This assumption works because the Compton wavelength of the chameleon particle is always much smaller than the width of the vacuum chamber walls for the parameter range of interest.

\begin{table}
	\begin{center}
  	\begin{tabular}{ | l | c | }
    		\hline
		{\bf Material}													&  	{\bf $\rho~(\mathrm{g} / \mathrm{cm}^3)$}		 \\ \hline
		source mass (aluminum)										&	2.7														\\
		vacuum	($6 \times 10^{-10}$ Torr)			&	$6.6 \times 10^{-17}$											\\
		vacuum chamber walls (steel)									&	7											\\
    		\hline
  	\end{tabular}
\end{center}
\caption{Densities of the materials in the experiment.}
\label{tab:densities}
\end{table}

The convergence time of this method is highly dependent upon the initial guess for the field configuration. There is a delicate tradeoff --- within dense regions ({\it i.e.}, source sphere and chamber walls), the equation of motion is highly nonlinear, and small steps are required to ensure convergence; within the vacuum region, on the other hand, the equation is approximately linear but can take many steps to reach the much larger field value. Steps small enough to ensure convergence in the dense regions make the convergence time in the vacuum region intolerably large, while steps large enough to converge inside the vacuum make the numerical scheme unstable in dense regions.

To address this issue we begin with a course-grained simulation, where $\phi$ in the dense areas is forced to minimize its effective potential as a boundary condition.  This is done only in regions where the Compton wavelength is more than an order of magnitude smaller than the grid spacing, so the chameleon is expected to minimize $V_{\rm eff}$ everywhere in the region.  The resulting course-grained output for $\phi$ is then interpolated into an initial guess for a higher resolution run. This method allows $\phi$ to quickly relax to its solution in the vacuum, while holding $\phi$ fixed in the numerically unstable regions.

\section{Successive Steps Towards Realistic Set-Up}
\label{sec:results}

In this Section we describe the results of the numerical integration, presented as successive steps towards the realistic experimental set-up. First, to make contact with our earlier analysis,
we use the 3D code to check the approximate analytical expression used in~\cite{Hamilton:2015zga} to place constraints on the chameleon parameter space. Remarkably, we find
only a $ 20 \%$ difference. As our next step, we compare the realistic cylindrical vacuum chamber to a spherical vacuum chamber of the same radius. This will determine
how sensitive the force calculation is to the ``spherical cow" approximation.  Here, we find an 18\% difference in the resulting acceleration at the interferometer between these two cases. Next
we examine the impact of offsetting the source mass from the center of the vacuum chamber, as is done in the actual experiment. We find the difference in acceleration at the location of the interferometer
to be negligible.  As our final step, we examine the effect of accounting for a circular bore through the source mass, as in the experiment.  Again, we find the difference in acceleration to be negligible.
For all the checks performed in this Section (except Sec.~\ref{comp1}), we assume $\Lambda = \Lambda_0 = 2.4$ meV, $M = 10^{-3} M_\mathrm{Pl}$, and focus on the power-law $n = 1$ following~\cite{Hamilton:2015zga}.

\begin{figure}
\centering
\includegraphics[width = 0.6 \linewidth]{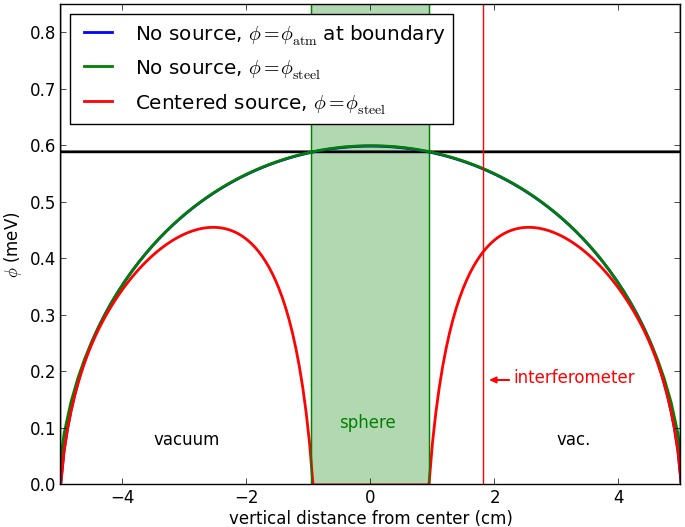}
\caption{The chameleon field as a function of distance along the center of the spherical vacuum chamber.  The black horizontal line marks the central value of $\phi$ predicted inside an empty chamber using~\eqref{phi_vac}. The red vertical line denotes the location of the interferometer.  We find essentially no difference between letting $\phi$ minimize its potential in atmosphere {\it vs} in steel at the walls.}
\label{fig:comparison}
\end{figure}

\subsection{Comparison to analytic approximation}
\label{comp1}

As a check on the code, we integrate the chameleon equation of motion under the same conditions as those explored in~\cite{Hamilton:2015zga}: a spherical vacuum chamber of radius $R_{\rm vac} = 5$~cm.
(As already mentioned, the actual vacuum chamber is not a sphere, and a better estimate for its effective radius is 6~cm, but for the purpose of comparing with earlier work we use the same parameters as~\cite{Hamilton:2015zga}.  This includes matching the parameters\footnote{The $\Lambda = 0.1$ meV  value is chosen solely for the purpose of comparison with the 1D numerical results of \cite{Hamilton:2015zga}. For the rest of our analysis we will use the fiducial dark energy value $\Lambda = \Lambda_0 = 2.4$ meV.}
$\Lambda = 0.1$ meV, $M = 10^{-3} M _\mathrm{Pl}$.) The field profile is calculated everywhere inside the chamber for 3 separate cases:

\begin{enumerate}

\item Without source mass ({\it i.e.}, empty vacuum chamber), and with boundary condition $\phi \rightarrow \phi_\mathrm{atm}$ at $r = R_{\rm vac}$.

\item Without source mass, and with boundary condition $\phi \to \phi_\mathrm{steel}$ at $r = R_{\rm vac}$.

\item Including a source mass of radius $r_{\rm s} = 1$~cm at the center of the chamber, imposing the same boundary condition as in Case~2.  

\end{enumerate}

\noindent The density of the different parts of the experiment are listed in Table~\ref{tab:densities}. (For Case 1, we use $\rho = 10^{-3}~{\rm g}/{\rm cm}^3$ for atmospheric density.)

The results are shown in Fig.~\ref{fig:comparison}. The chameleon field profiles in Cases 1 and 2 ({\it i.e.}, the cases without source mass), shown as the blue and green curves respectively, are virtually identical, leading us to conclude that the boundary conditions imposed at the vacuum chamber walls are unimportant to the dynamics near the center of the vacuum chamber. The black horizontal line indicates the central $\phi$ value predicted by~\eqref{phi_vac} with $\xi = 0.55$, as found in~\cite{Hamilton:2015zga}. We see that Cases 1 and 2 closely match this approximate constant solution near the center, in particular at the location of the interferometer (red vertical line), confirming that the code's results are consistent with~\cite{Hamilton:2015zga}.

\begin{figure}
  \begin{minipage}[b]{0.40\linewidth}
    \includegraphics[width=\linewidth]{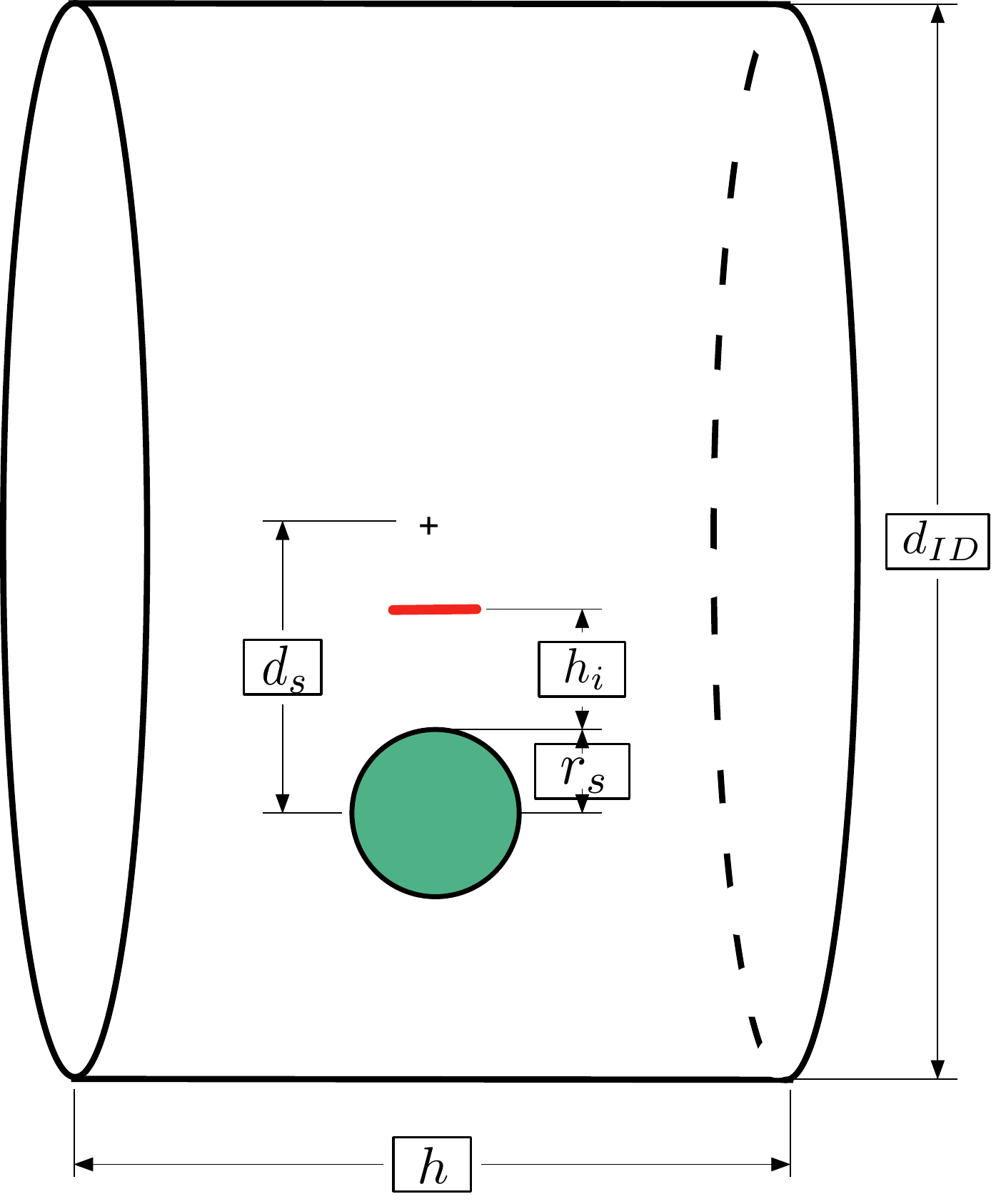}
  \end{minipage}%
  \begin{minipage}[b]{0.50\linewidth}
\begin{tabular}[b]{ | l | c | c |}
    		\hline
		Spherical source radius			&	$r_{\rm s}$	&	0.95 cm	\\ \hline
		Diameter of bore through source	&	$r_{\rm bore}$	&	0.30 cm	\\ \hline
		Location of spherical source		&	$d_{\rm s}$	&	2.55 cm 	\\ \hline
		Location of interferometer			&	$h_{\rm i}$	&	0.88 cm 	\\ \hline
		Inner diameter of vacuum chamber	&	$d_{\rm ID}$	&	12.2 cm	\\ \hline
		Vacuum chamber height			&	$h$   		&	7.1 cm	\\ \hline
  	\end{tabular}
	\vspace{3 mm}
\end{minipage}
\caption{{\bf Diagram and dimensions of experimental setup.}~ \small  The cross marks the center of the vacuum chamber.  The vacuum chamber walls are $\sim$ 2 cm thick, which is much greater than the Compton wavelength of the chameleon inside steel in all cases examined.}
\label{geometry}
\end{figure}

Case 3, shown as the red curve, includes the source mass and allows us to calculate the acceleration on a test atom exactly and directly
using~\eqref{acceleration}. The acceleration is attractive (pointing towards the center) near the source mass, but is repulsive
(pointing away from the center, and towards the chamber walls) further out. At the location of the interferometer,\footnote{The atoms actually traverse nearly 5 mm during the acceleration measurement.  Following~\cite{Hamilton:2015zga}, we approximate the atoms' average distance from the source mass as 8.8 mm.} the answer is
$a = 5.0 \times 10^{-10} ~ \mathrm{m} / \mathrm{s}^2$ towards the source mass. The value calculated in~\cite{Hamilton:2015zga}
using the approximate ``thin-shell" expression~\eqref{accel_approx} was $6.4 \times 10^{-10} ~ \mathrm{m} / \mathrm{s}^2$, an overestimate of approximately
$ 20 \%$. (As already mentioned, however, this is compensated by a slight underestimate of the vacuum chamber radius. The actual radius is 6~cm, resulting
in a larger acceleration at the location of the interferometer.) 

\subsection{Comparison: spherical {\it vs} cylindrical vacuum chamber}

\begin{figure}
\subfigure[Field profile]{%
\includegraphics[width=0.5\linewidth]{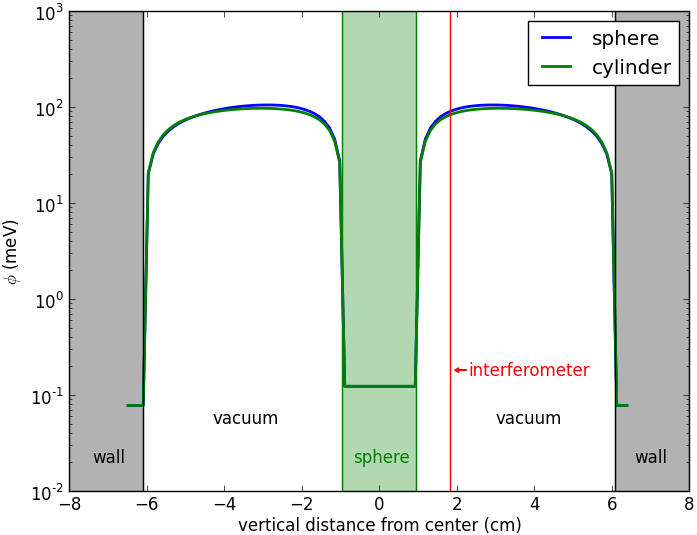}
\label{sph_cyl_prof}}
\subfigure[Acceleration]{%
\includegraphics[width=0.5\linewidth]{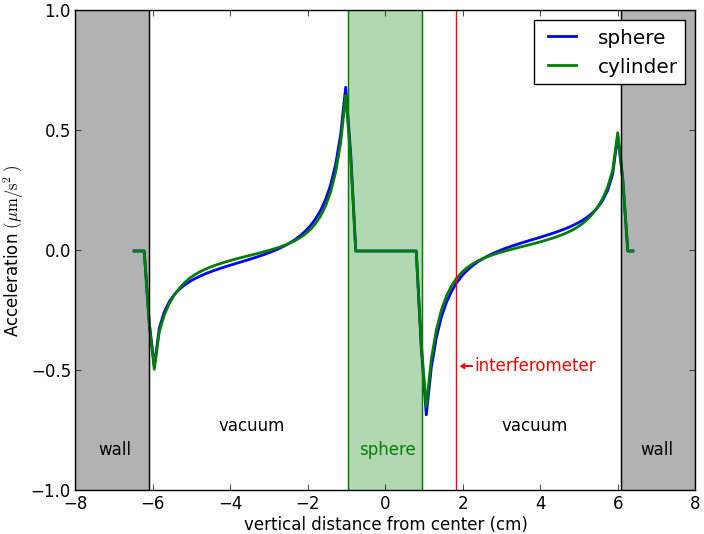}
\label{sph_cyl_accel}}
\caption{ {\bf Spherical {\it vs} cylindrical vacuum chamber.}~ \small Chameleon profile and acceleration as a function of distance from the center of the spherical source mass,
for a spherical (blue curve) and cylindrical (green curve) vacuum chamber. The dimensions of the cylindrical vacuum chamber are chosen to match that of the experiment in~\cite{Hamilton:2015zga} 
and are shown in Fig.~\ref{geometry}. The radius of the sphere is chosen to match the inner radius of the cylinder. At the location of the interferometer (red vertical line), the acceleration
in the spherical case is 18\% larger than in the cylindrical chamber.}
\label{sph_cyl}
\end{figure}

Next we examine the effect of approximating the cylindrical vacuum chamber as a sphere. For this purpose we assume a cylindrical geometry that matches the actual vacuum chamber
used in the experiment~\cite{Hamilton:2015zga}. As shown in Fig.~\ref{geometry} (except that the source mass in the present case is centered rather than offset), 
the vacuum chamber is a short cylinder, with inner radius of 6.1~cm, turned so that the axis of the cylinder is perpendicular to Earth's gravity. 
For comparison, we choose a sphere of the same radius, $R_{\rm vac} = 6.1$~cm, such that the distance between the source mass and the
vacuum chamber wall is the same in the direction of the interferometer. This makes for a fair comparison since, keeping the distance between the source mass and interferometer fixed, 
the chameleon gradient at the location of the interferometer is primarily influenced by its distance from the vacuum chamber wall~\cite{Upadhye:2012qu}.  Recall also that we are now going back to the cosmologically-motivated value of $\Lambda = \Lambda_0 = 2.4$ meV.

The results, shown in Fig.~\ref{sph_cyl}, demonstrate a minor departure between the cylinder {\it vs} the sphere. In particular, the acceleration at the interferometer is 18\% larger for the sphere than
for the cylinder.

\begin{figure}
\subfigure[Field profile]{%
\includegraphics[width=0.5\linewidth]{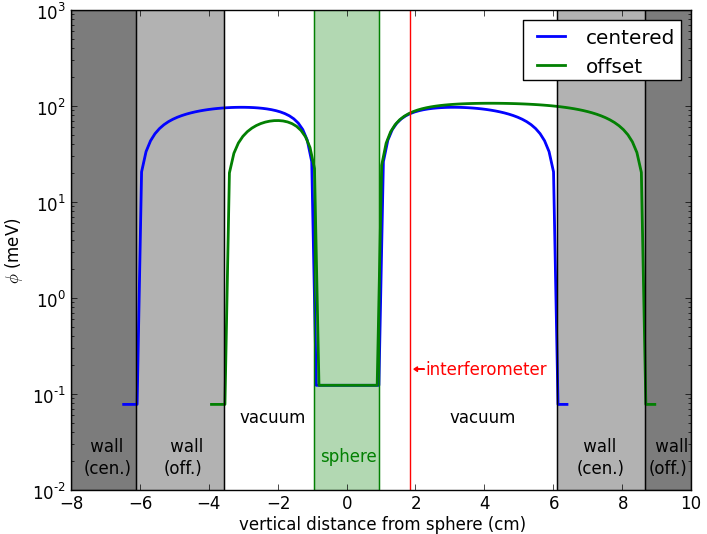}
\label{cen_off_prof}}
\subfigure[Acceleration]{%
\includegraphics[width=0.5\linewidth]{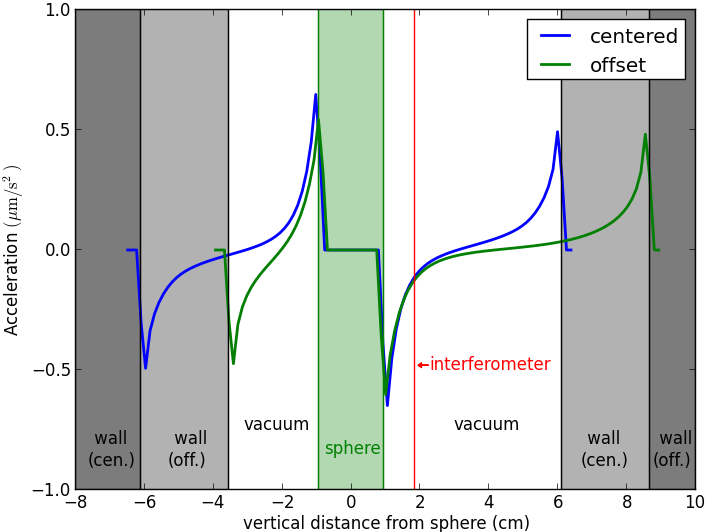}
\label{cen_off_accel}}
\caption{{\bf Source mass centered {\it vs} offset.}~ \small Same plot as the previous figure, now comparing a source mass at the center (blue curve) and offset by 2.55~cm from the center (green curve), as in the actual experiment. As in the previous figure, the dimensions of the cylindrical chamber match those of the experiment. Although the field profile is altered by the offset, the acceleration at the interferometer (red vertical line) changes by less than 1\%.}
\label{cen_off}
\end{figure}

\subsection{Comparison: source mass offset {\it vs} centered}

We now examine the effect of moving the source mass away from the center of the cylindrical vacuum chamber. For this purpose we once again assume a cylindrical geometry that matches the actual vacuum chamber used in the experiment~\cite{Hamilton:2015zga}, with dimensions listed in Fig.~\ref{geometry} (except with a solid source mass). We compare the chameleon profile and acceleration between a source mass at the center
and a source mass located 2.55~cm below the center, as in the actual experiment. The distance to the interferometer is kept fixed. The results, shown in Fig.~\ref{cen_off}, demonstrate that although the acceleration profiles are different in certain regions of the vacuum chamber, the difference at the interferometer is negligible. Had the interferometer been located further away from the source, the difference in acceleration would have been more significant.

\begin{figure}
\subfigure[Field profile]{%
\includegraphics[width=0.5\linewidth]{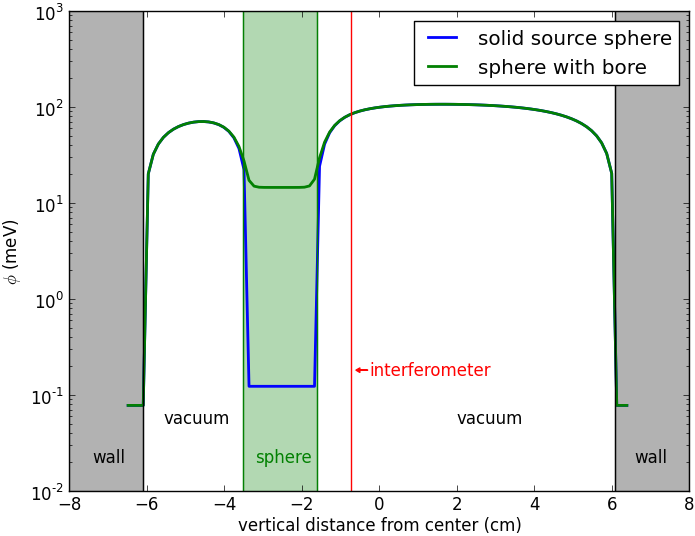}
\label{bore_prof}}
\subfigure[Acceleration]{%
\includegraphics[width=0.5\linewidth]{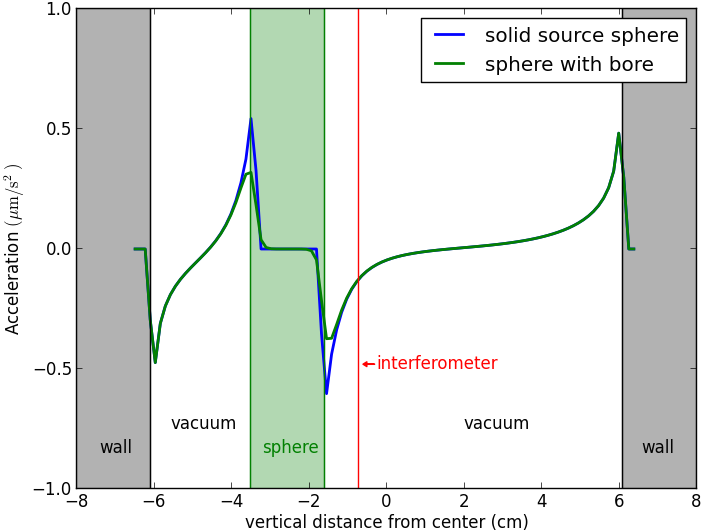}
\label{bore_accel}}
\caption{{\bf Source mass with {\it vs} without bore.}~ \small Same as the previous two figures, but now comparing a solid source mass (blue curve) against one with a 3~mm diameter circular bore through the center (green curve), as in the experiment. All other dimensions are chosen to match those of the experiment. The only significant difference is inside the sphere, as the green line passes through the center of the bore, so it is still in vacuum.  The acceleration at the interferometer (red vertical line) again changes by less than 1\%.}
\label{bore}
\end{figure}

\subsection{Comparison: solid source mass {\it vs} source mass with bore}

As a final step towards the experimentally realistic setup, we examine the effect of a vertical circular bore through the center of the spherical source mass.  We use the dimensions listed Fig.~\ref{geometry}, only in one case without the bore. The results, shown in Fig.~\ref{bore}, show that the difference in acceleration at the interferometer is again negligible. The difference is, however, significant within the source mass.  This is because the plot shows the chameleon profile through the center of the bore, a path which is in vacuum from wall to wall.  Indeed, when inside the sphere the bore acts as a miniature vacuum chamber, and the chameleon field reaches a value such that the Compton wavelength is comparable to the radius of the bore.

\section{Simulation of the Experiment}
\label{realsim}

We are now in position to simulate the experiment~\cite{Hamilton:2015zga} and derive realistic constraints on chameleon parameters. Once again the dimensions
of the vacuum chamber are sketched and listed in Fig.~\ref{geometry}. The material densities are listed in Table~\ref{tab:densities}. Following~\cite{Hamilton:2015zga}
and as assumed in the previous Section, we focus on the power-law $n=1$ and assume $\Lambda = \Lambda_0 = 2.4$~meV. 

The chameleon profiles are plotted in Fig.~\ref{exp}, for $M$ ranging from $10^{-5} M_\mathrm{Pl}$ to $M_\mathrm{Pl}$. The first thing to note from Fig.~\ref{exp} is that the field profile
inside the vacuum region is relatively insensitive to $M$. This can be understood as follows. On the one hand, in the vacuum region the density is effectively zero. Since $M$ only appears in the equation of motion as $\rho / M$, the chameleon equation of motion is essentially independent of $M$ in that region. The only dependence comes from the dense regions (source mass and chamber walls). But even so,
the chameleon is screened and  minimizes its effective potential at a very small field value in those dense regions, and for all intents and purposes $\phi \simeq 0$ there relative to the much larger field
value in the bulk of the chamber. This is why the profile is quite insensitive to $M$ inside the chamber. (For larger values of $M$ than considered here, the source mass and chamber walls eventually become unscreened and this argument would no longer hold.)

\begin{figure}
\centering
\includegraphics[width=0.7\linewidth]{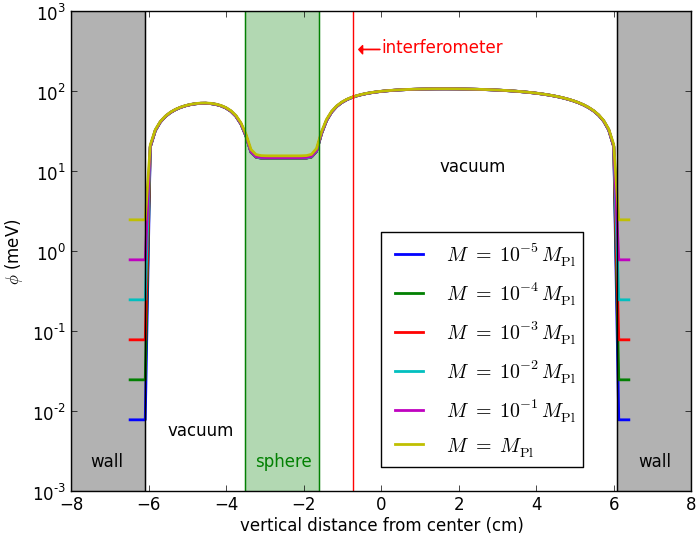}
\label{exp_prof}
\caption{\small Simulation of the experimental configuration, for values of $M$ ranging from $10^{-5} M_\mathrm{Pl}$ to $M_\mathrm{Pl}$.  We find that the profiles in vacuum are nearly identical, differing only in the walls.  The field values inside the metal of the source mass also scale with $M$, but we are showing a path that passes through the center of the bore in the source mass.  The bore acts as a miniature vacuum chamber, so instead the chameleon field goes to an $M$-independent value such that the Compton wavelength is of order the radius of the bore.}
\label{exp}
\end{figure}

The acceleration at the interferometer can be calculated using the gradient of the chameleon profiles. Since $\vec{\nabla}\phi$ at that position is essentially independent of $M$, the only dependence
on this parameter comes from the prefactor of $1/M$ in the expression~\eqref{acceleration} for the acceleration.  We find the resulting acceleration due to the chameleon field at the interferometer to be
\be
a = \frac{\vec \nabla \phi}{M} = 1.2 \times 10^{-4} ~\frac{M_\mathrm{Pl}}{M}~\mu \mathrm{m} / \mathrm{s}^2~.
\label{grad_phi}
\ee
As a particular example, with $M = 10^{-4} ~M_\mathrm{Pl}$ this yields an acceleration at the interferometer of $1.2 ~\mu \mathrm{m}/\mathrm{s}^2$.  The thin-shell approximate method used in~\cite{Hamilton:2015zga, Burrage:2015lya} yields an acceleration of $1.4 ~\mu \mathrm{m}/\mathrm{s}^2$, a difference of $\sim 20\%$.

The atom interferometry experiment~\cite{Hamilton:2015zga} placed an upper limit of $a < 5.5 ~ \mu \mathrm{m} / \mathrm{s}^2$ (95\% confidence level) on the chameleon acceleration. As can now be calculated from \eqref{grad_phi}, this corresponds to $M \leq 2.3 \times 10^{-5} M_\mathrm{Pl}$. Remarkably, this is the {\it same constraint} as quoted in~\cite{Hamilton:2015zga} using the approximations describe above. The reason for this coincidence is that these authors slightly {\it underestimated} the radius of the vacuum chamber (5~cm instead of the actual 6~cm), which just so happens to compensate the {\it overestimate} inherent in the approximate thin-shell method. 

\section{Forecasts for ongoing and upcoming experiments}
\label{sec:forecasts}

In this Section we describe two upcoming experiments that will place even tighter constraints on the chameleon theory's parameters.  The first is an improvement upon the experiment~\cite{Hamilton:2015zga}, performed by the same authors, and is currently underway.  The second is a proposed experiment for NASA's Cold Atom Laboratory~\cite{CAL} aboard the International Space Station.

\begin{figure}
\subfigure[Field profile]{%
\includegraphics[width=0.5\linewidth]{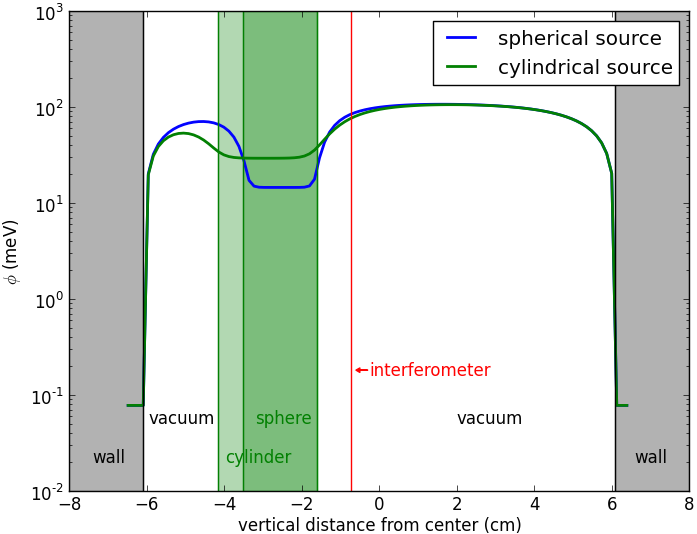}
\label{cyl_src_prof}}
\subfigure[Acceleration]{%
\includegraphics[width=0.5\linewidth]{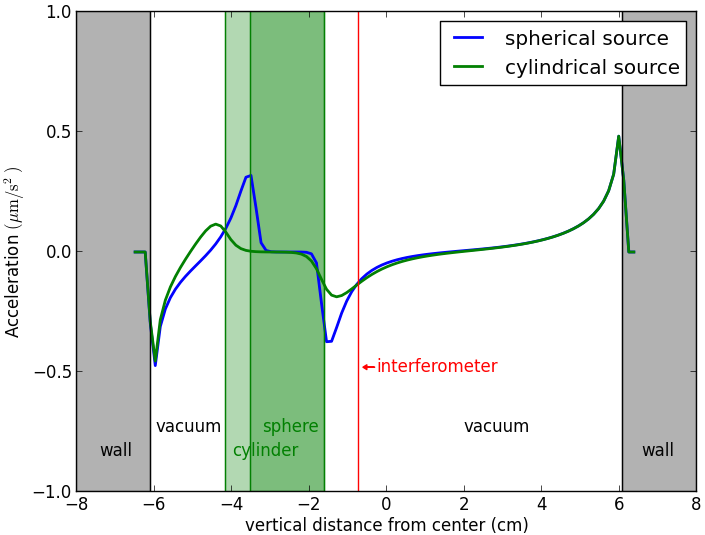}
\label{cyl_src_accel}}
\caption{{\bf Spherical source mass {\it vs} cylindrical source.}~ \small Comparison between two experimental setups: that of~\cite{Hamilton:2015zga} (blue line) and of an improved version of the experiment that is currently underway (green line).  The main difference is that the latter employs a tungsten cylinder as the source mass, while the former used an aluminum sphere.  The cylinder has a wedge cut out of it, allowing for vastly improved control over systematics.  These show that the cutout comes at no cost to the chameleon signal, in fact, the cylinder confers a 5\% stronger chameleon force over the previous setup. }
\label{cyl_src}
\end{figure}

\subsection{Laboratory experiment: spherical source with wedge}
This experiment is similar to~\cite{Hamilton:2015zga}, except with greater sensitivity thanks to a variety of technical improvements such as colder atoms, additional vibration isolation, and the atoms are now launched upwards (rather than dropped) to allow them to spend more time near the source mass.  Another key difference is that the source mass is now a tungsten hollow cylinder with a wedge cutout.  This geometry was chosen so that the source mass may be moved away from the interferometer without breaking the atom/laser beam line, allowing for better control of the systematic errors.

To evaluate the sensitivity of this new setup, we perform a comparison against the geometry described in the previous Section.  As before, we assume $\Lambda = 2.4$ meV, $M = 10^{-3} M_\mathrm{Pl}$, and $n = 1$.  The source mass is a hollow cylinder with an outer diameter of 2.54~cm, inner diameter 0.99~cm, and length 2.56~cm. It is made of tungsten, which has a density of 19.25 $\mathrm{g} / \mathrm{cm}^3$~. There is a wedge cut out of one side with thickness 0.50~cm. 

The results, plotted in Fig.~\ref{cyl_src}, show that the new setup produces an acceleration that is 5\% larger than the previous one.  This comes with a large improvement in systematic errors as well, which will allow for much greater sensitivity.  Altogether, the new setup is expected to improve upon the limit of $M\leq 2.3\times 10^{-5} M_{\rm Pl}$ from~\cite{Hamilton:2015zga} by 1-2 orders of magnitude.

\subsection{Space-based experiment: Cold Atom Laboratory}

\begin{figure}
\centering
\includegraphics[width=0.7\linewidth]{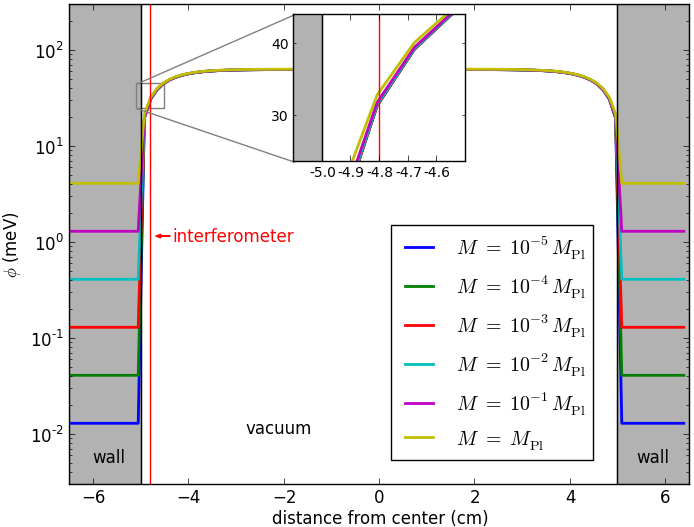}
\label{CAL_prof}
\caption{\small Same plot as Fig.~\ref{exp}, but for the empty rectangular vacuum chamber of the CAL experiment.  The field profiles are taken along the long axis of the vacuum chamber. Again, we find that the profiles in vacuum are nearly identical.}
\label{CAL}
\end{figure}

This experiment is proposed to take place inside NASA's Cold Atom Laboratory~\cite{CAL} aboard the International Space Station, and is currently scheduled to be launched in 2017.  Ground-based experiments are limited in that Earth's gravity causes the atoms to only spend a limited amount of time near the source mass.  Performing the experiment in space obviates this issue, allowing for greater sensitivity.\footnote{Long interaction times may also be achieved in ground-based experiments by dropping both the source mass and the atoms, such as in an Einstein elevator~\cite{Lotz:2014} or in a zero-gravity flight~\cite{Geiger:2011}.}

The experiment consists of an empty rectangular vacuum chamber with $3 \times 3$ cm cross section and length 10 cm.  Interferometry is performed with atoms located on an axis parallel, and close, to the central long axis of the vacuum chamber.  The atoms' acceleration may be measured anywhere along this path (up to within $\sim$ 0.5 mm of the walls).  The walls are made of glass, with a density of roughly 2.5~$\mathrm{g} / \mathrm{cm}^3$. We assume the same chameleon parameters as in the previous Section.

The resulting chameleon field profiles along the long axis of the vacuum chamber are shown in Fig.~\ref{CAL}.  If the measurement is performed 2~mm from the vacuum chamber walls, we find an acceleration
\be
a = 2.7 \times 10^{-3}~\frac{M_\mathrm{Pl}}{M} ~\mu \mathrm{m} / \mathrm{s}^2~,
\ee
towards the wall.  This value is independent of $M$ (to within 5\%) as long as $M \lesssim M_\mathrm{Pl}$.

This demonstrates that, thanks to how close the atoms may get to the walls, the magnitude of the chameleonic acceleration is similar to that of the ground-based experiments.
This result, combined with the much longer interaction times between the source and the atoms, as well as common-mode rejection of the influence of vibrations achieved by running two simultaneous atom interferometers with potassium and rubidium atoms, respectively, gives hope for much tighter restrictions on chameleon parameter space. 
%An improvement by 1-3 orders of magnitude relative to~\cite{Hamilton:2015zga} might be possible. 
An optimized version could in principle be designed to be sensitive to the entire parameter space $M\lesssim M_{\rm Pl}$.

\section{Conclusions}
\label{sec:conclusions}

In this paper we have, for the first time, solved the three-dimensional nonlinear PDE governing the chameleon scalar field inside a vacuum chamber, for static configurations. Along the way, through a series of increasingly
realistic runs, we have explored the impact of various approximations made in earlier work. In particular, approximating the cylindrical vacuum chamber with a sphere while keeping the
distance between the interferometer and the nearest chamber wall fixed, results in an 18\% difference in acceleration at the location of the interferometer. Moving the source mass
away from the center while keeping the distance to the interferometer fixed, has negligible effect on the measured acceleration. We then solved for the chameleon field in an experimentally realistic setup for $10^{-5} M_\mathrm{Pl} \leq M \leq M_\mathrm{Pl}$, finding that the chameleon profile is largely independent of $M$ inside the vacuum chamber. We have ruled out $M <  2.3 \times 10^{-5} M_\mathrm{Pl}$ at the 95\% confidence level for $n = 1$ and $\Lambda = \Lambda_0$, based on the upper bound on the acceleration reported in~\cite{Hamilton:2015zga}.   Finally, we have performed a preliminary analysis for upcoming experiments which can, in principle, sense the entire parameter space $M \lesssim M_{\rm Pl}$.

In the future it will be interesting to use the techniques described here to explore the effects of different source mass geometries, as it may be possible to optimize experiments for greater sensitivity.  Additionally, as experimental results become more precise, so too should the theoretical predictions.  This may necessitate more accurate modeling of the vacuum chamber geometry. Our method may also prove to be an invaluable tool for such a purpose.\\

\noindent {\bf Acknowledgements:} We are grateful for helpful discussions with Lasha Berezhiani, S\'{e}bastien Clesse, Rehan Deen, Andrei Ivanov, Sandrine Schl\"{o}gel, Amol Upadhye, and Nan Yu.
B.E. and J.K. are supported in part by NSF CAREER Award PHY-1145525, NASA ATP grant NNX11AI95G and a New Initiative Research Grant from the
Charles E. Kaufman fund of The Pittsburgh Foundation. H.W. is supported by the David and Lucile Packard Foundation, the DARPA Young Faculty Award N66001-12-1-4232, NSF grant PHY-1404566, and NASA grants NNH13ZTT002N, NNH13ZTT002N, and NNH11ZTT001N. P.H. thanks the Austrian Science Fund (FWF): J3680.

\section*{Appendix: Numerical Algorithm}

In this Appendix we offer some details on the numerical approach used to integrate the chameleon equation of motion~\eqref{eom}. This equation is a non-linear 
Poisson-Boltzmann equation of the form:
\be
\nabla^2 \phi = \rho(x, \phi)~.
\ee
Let us illustrate the method with the simplest case of one spatial dimension. In that case the Laplacian operator on the left-hand side with a finite
difference operator~\cite{NR}
\be
\frac{1}{ ( \Delta x)^2} \bigg(\phi(x + \Delta x)   - 2 \phi(x) +  \phi(x - \Delta x)  \bigg) = \rho(x, \phi)~.
\ee
This approximation follows from the second-order Taylor expansion of $\phi$, and becomes exact as $\Delta x \to 0$ for smooth functions.
Isolating $\phi(x)$ gives a relation that may be used to iteratively solve for $\phi$:
\be
\phi(x) = \frac{1}{2} \bigg( \phi(x + \Delta x) + \phi(x - \Delta x) -  (\Delta x)^2 \rho(x, \phi) \bigg)~.
\ee
To use this equation, we begin with an initial guess for $\phi(x)$ and apply this equation at each point successively from one edge of the integration to the other.  This process is repeated iteratively until $\phi(x)$ converges on a solution.  If the neighboring $\phi$ values on the right-hand side come from the previous iteration, this is known as the Jacobi method.  Using the most recently computed value of $\phi$ on the right-hand side converges more quickly and is known as the Gauss-Seidel method. We follow the latter method in our numerical integration. 

This process generalizes straightforwardly to three dimensions. Here the finite difference expression becomes
\begin{align} \nonumber
\phi(x, y, z) = \frac{1}{6} \bigg( &\phi(x + h, y, z) + \phi(x - h, y, z) \\ \nonumber
					      &\phi(x, y + h, z) + \phi(x, y - h, z) \\
					      &\phi(x, y, z + h) + \phi(x, y, z - h) - h^2 \rho(x, y, z, \phi) \bigg)~,
\end{align}
where $h$ is the grid spacing. Care must be taken at the edges. In this case we replace any occurrence of the type $\phi(x, -h, z)$ with $\phi(x, h, z)$.  This effectively imposes the
boundary condition that the normal derivative of $\phi$ vanish at the edge of the simulation.

Depending on the form of $\rho$, this algorithm may converge very slowly, or it may not converge at all.  We can cure such speed/stability issues by introducing an over/under correction scheme:
\be
\phi^{(n+1)}(x) = \phi^{(n)} - \alpha \left(\phi^* - \phi^{(n)}\right)~.
\label{relaxation}
\ee
Here, $\phi^{(i)}$ represents the $i$-th iteration of $\phi$, and $\phi^*$ is predicted by Gauss-Seidel based on the previous iteration. Meanwhile, $\alpha$ is the relaxation parameter and can take any value in the interval $0<\alpha< 2$.  For $0 < \alpha < 1$, the algorithm converges more slowly than Gauss-Seidel, but allows for numerical instabilities to be tamed.  For $\alpha = 1$ the right-hand side reduces to $\phi^*$, hence the method reduces to Gauss-Seidel.  If $1 < \alpha < 2$, the method will converge more quickly, but is also more likely to be unstable. Due to the non-linear nature of the chameleon equation, we encountered significant numerical instabilities, especially in the dense regions. This was cured by taking $\alpha < 1$.

\end{document}